\documentclass[floatfix,%
reprint,%
superscriptaddress,%
amsmath,amssymb,%
aps,%
prb%
]{revtex4-2}

\usepackage{graphicx}
\usepackage{dcolumn}
\usepackage{bm}
\usepackage{physics}
\usepackage{hyperref}
\usepackage{xparse}
\usepackage{color}
\usepackage{subfigure}
\usepackage{longtable}
\usepackage[verbose]{placeins} 

\usepackage{float} 

\NewDocumentCommand{\rnl}{O{r} O{n} O{l}}{X_{#2 #3} (#1)}

\newcommand{\s}{_\mathrm{{\scriptscriptstyle S}}}

\newcommand{\x}{_\mathrm{{\scriptscriptstyle X}}}
\newcommand{\xc}{_\mathrm{{\scriptscriptstyle XC}}}

\newcommand{\F}{_\mathrm{{\scriptscriptstyle F}}}
\newcommand{\B}{_\mathrm{{\scriptscriptstyle B}}}
\newcommand{\unif}{\mathrm{{unif}}}

\newcommand{\PBE}{^\mathrm{{PBE}}}
\newcommand{\tPBE}{^\mathrm{{tPBE}}}

\begin{document}

\title{Thermal PBE in warm dense matter:  Does it matter and is it accurate?}

\author{Kushal Ramakrishna}
\email{k.ramakrishna@hzdr.de}
\affiliation{Information Services and Computing, Helmholtz--Zentrum Dresden--Rossendorf (HZDR), 01328 Dresden, Germany}

\author{Mani Lokamani}
\affiliation{Information Services and Computing, Helmholtz--Zentrum Dresden--Rossendorf (HZDR), 01328 Dresden, Germany}

\author{Zhandos A.~Moldabekov}
\affiliation{Institute of Radiation Physics, Helmholtz--Zentrum Dresden--Rossendorf (HZDR), 01328 Dresden, Germany}

\author{Tobias Dornheim}
\affiliation{Institute of Radiation Physics, Helmholtz--Zentrum Dresden--Rossendorf (HZDR), 01328 Dresden, Germany}
\affiliation{Center for Advanced Systems Understanding (CASUS), Helmholtz--Zentrum Dresden--Rossendorf (HZDR), 02826 G\"orlitz, Germany}

\author{Kieron Burke}
\affiliation{Departments of Chemistry and Physics, University of California, Irvine, USA}

\author{Attila Cangi}
\email{a.cangi@hzdr.de}
\affiliation{Center for Advanced Systems Understanding (CASUS), Helmholtz--Zentrum Dresden--Rossendorf (HZDR), 02826 G\"orlitz, Germany}

\date{\today}

\begin{abstract}
Conditional probability density functional theory has recently been used to derive the temperature dependence of the Perdew–Burke–Ernzerhof (PBE) generalized gradient approximation (GGA) for the exchange--correlation (XC) free energy.  We implement and systematically benchmark thermal PBE within Kohn–Sham density functional theory calculations of warm dense matter. Comparisons with the local density approximation (LDA) and PBE functionals, as well as thermal LDA, show that thermal PBE significantly improves the description of warm dense matter properties, including energies, forces, pressures, and electronic charge densities. In particular, thermal PBE exhibits close agreement with path integral Monte Carlo (PIMC) reference data at negligible additional computational cost. This work demonstrates the practical utility of thermal PBE as an accurate semilocal functional for simulations in the warm dense regime.
\end{abstract}

\maketitle

\section{Introduction}

Warm dense matter (WDM) is a state of matter defined by its unique combination of high temperature and density~\cite{vorberger2025roadmapwarmdensematter,wdm_book}. Specifically, it refers to a material heated to temperatures on the order of thousands to hundreds of thousands of Kelvin while remaining compressed to densities similar to those found in solids or liquids~\cite{bonitz2020ab,bonitz2024principles,doi:10.1063/5.0138955}. This regime is of particular interest because it is difficult to study experimentally but relevant to various astrophysical phenomena, such as the interiors of giant planets and white dwarf stars~\cite{saumon1,Militzer_2008,militzer1,nettelmann2011thermal,kritcher2020measurement}. From a technological point of view, understanding the properties of WDM is highly relevant for inertial confinement fusion~\cite{betti2016inertial,vorberger2025roadmapwarmdensematter} as both the fusion fuel and the ablator material have to pass through warm dense conditions on their way from room temperature to ignition. The study of WDM requires advanced experimental techniques, such as high--energy lasers, to create and probe these extreme conditions~\cite{zastrau_resolving_2014,Fletcher2015,Tschentscher_2017,falk_wdm}.
Because of the experimental challenges, researchers also rely on sophisticated computational techniques~\cite{doi:10.1063/5.0138955}.

Theoretical methods such as path integral Monte Carlo (PIMC)~\cite{PhysRevLett.121.255001,review,PhysRevLett.117.115701,Brown_PRL,dornheim2019static,PhysRevLett.129.066402,Dornheim2025} and finite--temperature Kohn--Sham density functional theory (DFT)~\cite{hohenberg1964inhomogeneous,mermin,kohn1965self} coupled to molecular dynamics (MD) are currently the most widely used methods for computing the properties of WDM. 
In the Mermin generalization of the Hohenberg--Kohn theorem to finite temperatures, the XC free energy is known to become temperature dependent~\cite{mermin}. In the past, many such DFT simulations used the local density approximation (LDA), for which the temperature dependence has been extensively derived, largely via quantum Monte Carlo (QMC) calculations~\cite{PhysRevLett.45.566,Brown_PRL,ksdt,PhysRevLett.117.156403,PhysRevLett.119.135001}.

However, modern DFT simulations of WDM almost always employ the PBE generalized gradient approximation, because of its higher accuracy in predicting energetic differences~\cite{10.1063/5.0315749}. Many such calculations are qualitatively successful, and often semi--quantitative in their accuracy, while ignoring the temperature--dependence in XC. This success could be explained by their relative accuracy in both the zero--{\em} and high--temperature limits (in the latter case, XC becomes relatively negligible~\cite{PhysRevB.93.245131}), and a study of this effect~\cite{PhysRevE.93.063207} showed a maximum relative error in the dc electrical conductivity of no more than 15\%. On the other hand, such effects might be crucial in a simulation, and other properties might show larger deviations. These include electronic structure~\cite{PhysRevE.93.063207,karasiev2018nonempirical,PhysRevB.101.195129,g1sl-72nz,PhysRevB.101.245141}, optical properties such as reflectivity~\cite{PhysRevB.99.214110} and electrical conductivity~\cite{PhysRevB.83.235120,ramakrishna2024electricalconductivitywarmdense}, as well as thermodynamic properties such as phase transitions and equation of state (EOS)~\cite{PhysRevB.99.214110,bonitz2024principles}. This represents a recognized source of systematic uncertainty in such simulations.
 
In this work, we investigate the importance of explicit temperature dependence in the XC free energy by evaluating the newly developed thermal PBE functional based on the concept of conditional probability density functional theory~\cite{kozlowski2023generalized}. 

We first summarize the recently derived thermal GGA and then provide a systematic assessment of the thermal PBE functional for various properties of warm dense hydrogen, a fundamental system for understanding WDM. This includes comparisons of total free energies, forces, and pressure (Fig.~\ref{forces_H_rs4}) and the accuracy of the electronic density with PIMC calculations (Fig.~\ref{fig:elec_dens}). We also provide a numerical assessment of our implementation relative to the LDA, PBE, and thermal LDA for the uniform electron gas in the relevant range of temperatures and densities (Figs.~\ref{xc_HEG}, and ~\ref{xc_HEG_pol}).
Furthermore, we provide additional comparisons of the thermal PBE functional with LDA, PBE, and thermal LDA for warm dense hydrogen, including the behavior of the total and XC free energies as a function of temperature (Figs.~\ref{hydrogen_rs4_dft_functionals}, and ~\ref{hydrogen_rs2_spinpol}), as well as comparisons of spatially resolved quantities such as the charge density, the electron localization function (ELF), and the density gradient (Figs.~\ref{fig:supp_tlda}, and \ref{fig:supp_rel_rel_elec_dens}).

\section{Results and discussion}
\label{ref_results}

We begin by summarizing the form of the thermal PBE (tPBE) XC free energy.  Within this framework, the XC free energy is expressed as
\begin{equation}
A\xc\tPBE(r\s,\zeta, s, \theta) = \int d^3r\ n(r)\, \epsilon\x^\unif(r\s)\, F\xc\tPBE(r\s,\zeta, s, \theta)\,, 
\label{eq.tpbe.Fxc}  
\end{equation}
where $\epsilon\x^\unif=-3k\F/(4\pi)$ is the exchange energy per particle of the uniform electron gas, $k\F=\left(3\pi^2n(r)\right)^{1/3}$ the Fermi wavevector, $F\xc\tPBE$ the enhancement factor of the thermal PBE functional, $r\s=\left(3/4 \pi n(r)\right)^{1/3}$ the Wigner--Seitz radius, $\zeta=(n{_\uparrow}(r)-n{_\downarrow}(r))/n(r)$ the relative spin orientation defined in terms of spin--up and spin--down densities, $s=|\nabla n(r)|/\left(2k\F n(r)\right)$ the dimensionless density gradient, and $\theta = T/T\F$ the reduced temperature with $T$ the electronic temperature and $T\F=k\F^2/(2k_B)$ the Fermi temperature, where  $k\B$ is the Boltzmann constant.  

The particular form of the thermal PBE enhancement factor, recently constructed to be consistent with
the conditional probability density functional theory calculations~\cite{kozlowski2023generalized}, is given by 
\begin{equation}
F\xc^{\tPBE}(r\s,\zeta, s, \theta) = \frac{F\xc^{\unif}(r\s, \zeta, \theta)}{F\xc^{\unif}(r\s, \zeta)}\,F\xc^{\PBE}(r\s,\zeta,s)\,,
\label{enhancement_factor_eqn}
\end{equation}
where $F\xc^{\unif}(r\s, \zeta, \theta)$ denotes the enhancement factor of the uniform electron gas at finite--temperature and $F\xc^{\unif}(r\s, \zeta)$ denotes that in the ground state.
Importantly, the temperature dependence of the thermal PBE functional enters solely through a thermal prefactor, defined as the ratio of enhancement factors for the thermal and ground--state uniform electron gas. This prefactor multiplies the ground--state PBE enhancement factor (which is independent of temperature) while containing all density gradient dependencies. As a result, thermal PBE consistently incorporates spin--polarization effects at finite--temperature, in contrast to other thermal GGA functionals where such spin dependence is absent~\cite{karasiev2018nonempirical}. Consequently, the resulting XC free energy of thermal PBE reduces to PBE at zero temperature and to finite--temperature LDA in the absence of density gradients. The thermal prefactor in Eq.~\ref{enhancement_factor_eqn} is further analyzed and illustrated in Appendix~\ref{appendix.thermal.prefactor}.

Accurate parameterizations for the ground--state XC energy of the uniform electron gas have been established for some time~\cite{perdew_zunger, perdew_wang,vwn}. More recently, the XC energy of the uniform electron gas at finite--temperature has also been parameterized~\cite{PhysRevLett.117.115701,PhysRevLett.119.135001}. The free energy per particle in this case is parameterized by
\begin{equation}
a\xc^\unif(r\s, \theta) = -\frac{1}{ r\s } \frac{ a(\theta) + b(\theta) r\s^{1/2} + c(\theta) r\s }{ 1 + d(\theta) r\s^{1/2} + e(\theta)r\s }\,,
\label{fxc_particle} 
\end{equation}
where $a(\theta)$, $b(\theta)$, $c(\theta)$, $d(\theta)$ and $e(\theta)$ are functions of reduced temperature (see Refs.~\cite{PhysRevLett.119.135001,PhysRevLett.117.115701} for parameterization). 

We have implemented the thermal PBE functional (as defined in Eq.~\ref{enhancement_factor_eqn}) in  LIBXC~\cite{lehtola2018recent,MARQUES20122272}, incurring negligible additional computational overhead compared to ground--state PBE. We validate our implementation by performing calculations for the uniform electron gas across a range of relevant temperatures and densities in Appendix~\ref{appendix.implementation}.

Having validated thermal PBE for the uniform electron gas, we next assess its performance for warm dense hydrogen --- a prototypical system that exhibits strong thermal effects, and pronounced density gradients. Hydrogen is not only of fundamental interest in itself but also serves as a benchmark system for EOS calculations under extreme conditions.

\begin{figure*}[!htbp]   
\centering    
\includegraphics[width=2.0\columnwidth]
{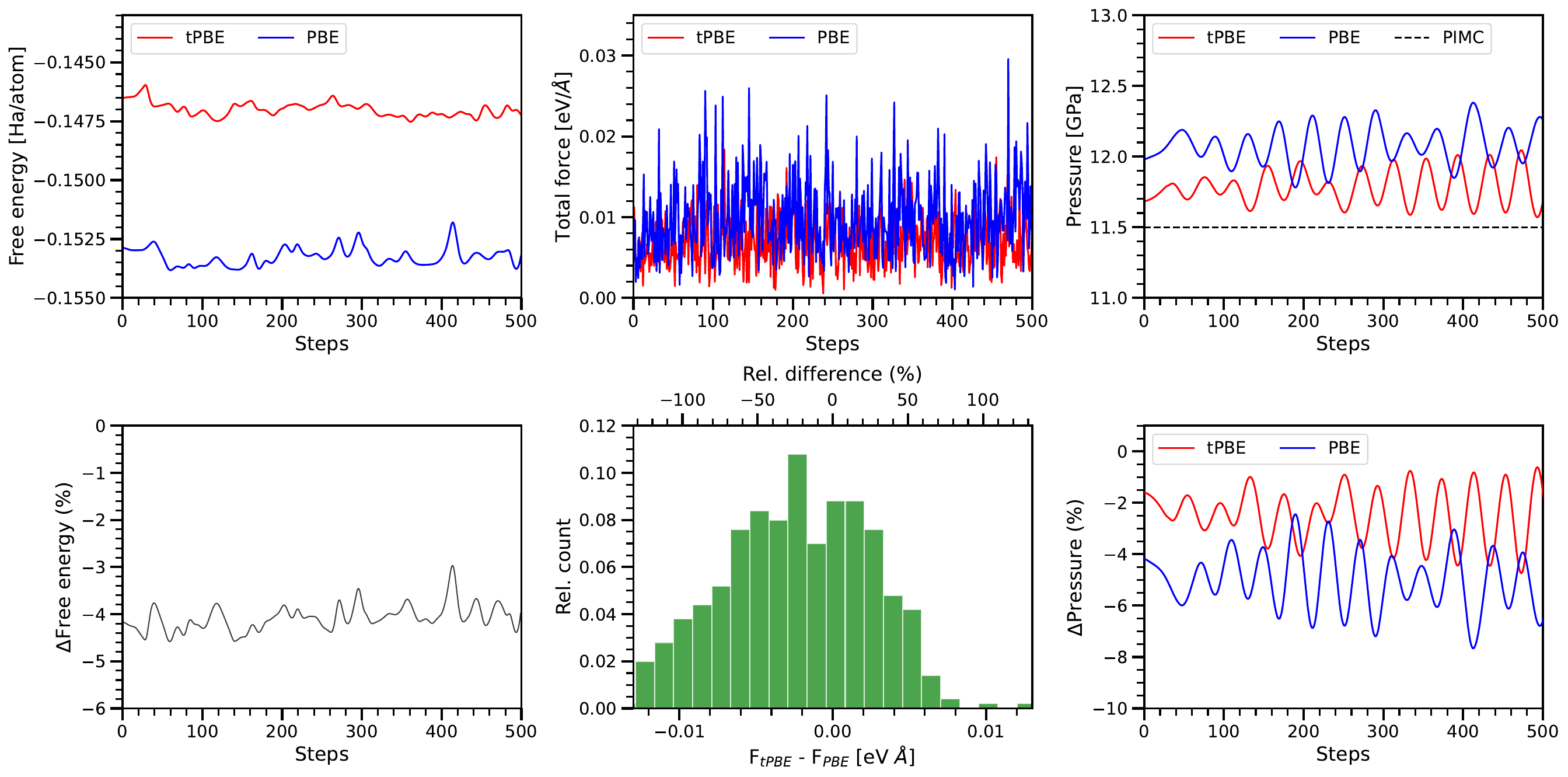} 
\caption{DFT--MD simulation of hydrogen at $r\s$=4 ($\rho$=0.042~g/cm$^{3}$), $T$=31,250~K ($\theta$=0.86). The plots in the top row show the free energy, force, and pressure per MD step using thermal PBE (red) and PBE (blue). The left plot in the bottom row shows the relative difference in free energy between thermal PBE and PBE. The center plot shows the histogram of instantaneous force differences $F_{\mathrm{tPBE}} - F_{\mathrm{PBE}}$, illustrating the distribution and typical magnitude of deviations between thermal PBE and PBE. The right plot in the bottom row shows the relative pressure difference of thermal PBE and PBE in comparison to PIMC~\cite{hu2011first}.}
\label{forces_H_rs4}   
\end{figure*}  

We first assess the thermal PBE functional in the context of DFT--MD simulations. To this end, we calculate the free energy, total force and pressure of hydrogen at $r\s=4$, $T$=31,250~K ($\theta$=0.86). The calculations are performed using the Vienna ab initio Simulation Package (VASP)~\cite{PhysRevB.47.558,PhysRevB.59.1758,KRESSE199615,PhysRevB.54.11169} using a Nose--Hoover thermostat~\cite{PhysRevA.31.1695} in the NVT ensemble with an ionic time--step of $\Delta t = 0.08$~fs. All calculations are performed in the spin--unpolarized configuration, corresponding to $\zeta = 0$. At WDM conditions ($T \sim 10{,}000$--$100{,}000$~K), hydrogen is fully dissociated and the electrons are delocalized, so the net spin polarization vanishes self-consistently; the system is genuinely nonmagnetic. The simulation cell contains 256 hydrogen atoms with 4,056 electronic bands included to ensure that the highest–band occupation remains below $\sim10^{-5}$. Brillouin zone sampling is carried out at the Baldereschi mean value point~\cite{PhysRevB.7.5212}. The top row of Fig.~\ref{forces_H_rs4} shows the time evolution of the free energy, total force, and pressure obtained with thermal PBE and ground--state PBE. Although both functionals produce stable trajectories, systematic differences are evident. 
The plots in the bottom row show the corresponding relative differences. The left plot indicates free energy relative differences between thermal PBE and PBE in the range 3--5\%, while the middle plot shows the histogram of instantaneous force differences $F_{\mathrm{tPBE}} - F_{\mathrm{PBE}}$. We observe a peaked distribution of force differences, with most samples concentrated around $\pm 0.01~\mathrm{eV / \AA}$, indicating small yet consistent deviations between thermal PBE and PBE forces. Pressure provides an especially stringent test, as it directly impacts EOS and Hugoniot predictions. The right plot in the bottom row compares results with PIMC reference data. The comparison reveals that thermal PBE yields pressures in close agreement with PIMC, with errors below 2\%, whereas ground--state PBE exhibits substantially larger deviations. These results are consistent with earlier studies demonstrating the critical role of thermal XC contributions for accurately describing thermodynamic properties of warm dense matter~\cite{PhysRevE.93.063207,karasiev2018nonempirical,PhysRevB.101.245141,bonitz2020ab,PhysRevB.101.195129,PhysRevB.99.214110}.

\begin{figure*}[!htbp]
    \centering
    \subfigure[]{\includegraphics[width=0.5\columnwidth]{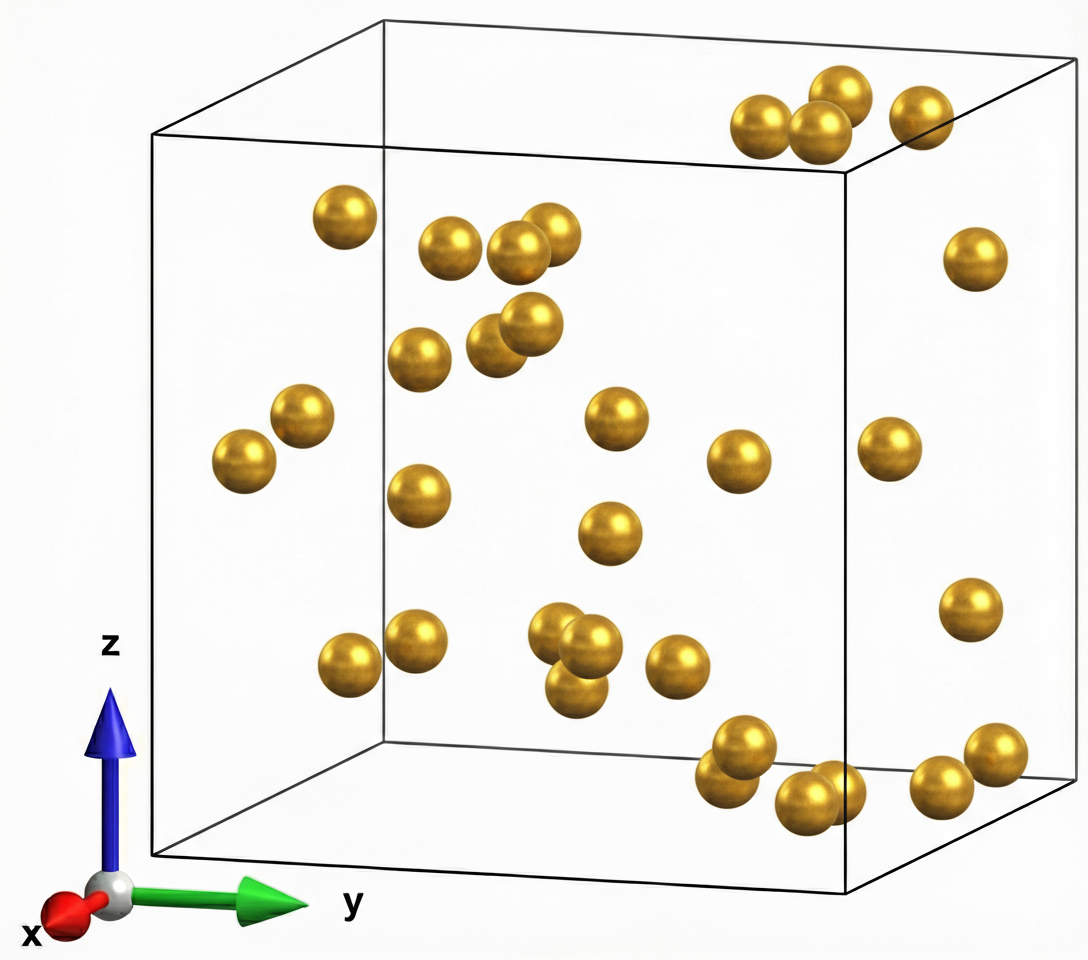}\label{fig:elec_dens_a}}\\[0.05em]
    \subfigure[]{\includegraphics[width=1.3\columnwidth]{charge_density_difference_planes_percentage_tpbe_pbe.png}\label{fig:elec_dens_b}}\\[0.05em]
    \subfigure[]{\includegraphics[width=1.3\columnwidth]{ELF_difference_planes_percentage_tpbe_pbe.png}\label{fig:elec_dens_c}}\\[0.05em]
    \subfigure[]{\includegraphics[width=1.3\columnwidth]{density_gradient_difference_planes_percentage_tpbe_pbe.png}\label{fig:elec_dens_d}}\\[0.05em]
    \subfigure[]{\includegraphics[width=1.6\columnwidth]{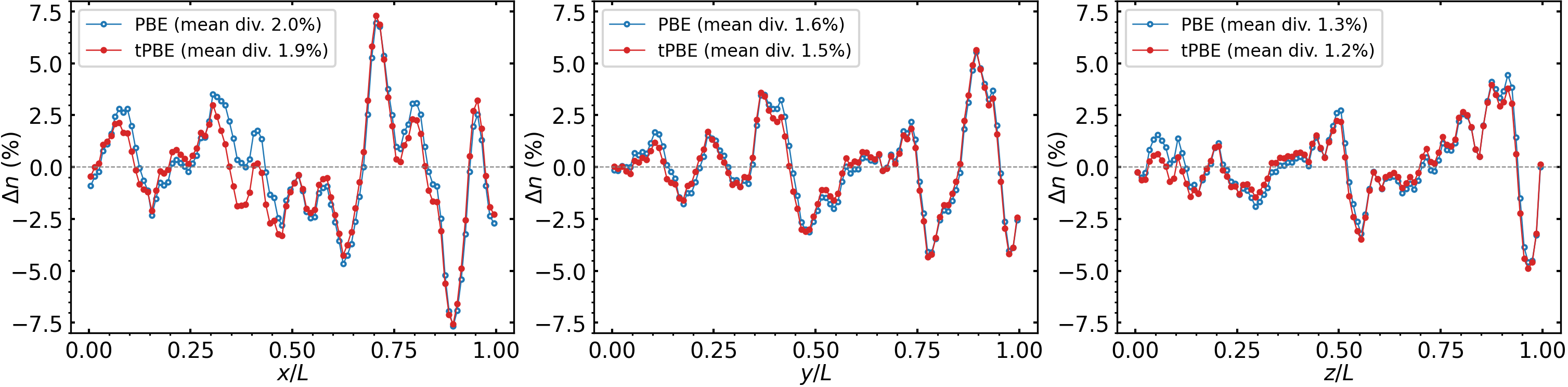}\label{fig:elec_dens_e}}
    \caption{Real--space electronic structure analysis of hydrogen at $r\s=4$, $\theta=1$. 
    (a) Simulation cell with ionic positions (gold spheres). 
    (b) Relative differences in electronic charge density, $(n_{\mathrm{tPBE}} - n_{\mathrm{PBE}})/n_{\mathrm{tPBE}}$, 
    along [001], [010], and [100] planes. 
    (c) Relative differences in ELF, 
    $(\mathrm{ELF}_{\mathrm{tPBE}} - \mathrm{ELF}_{\mathrm{PBE}})/\mathrm{ELF}_{\mathrm{tPBE}}$.
    (d) Relative differences in density gradient magnitude ($|\nabla n|$), $(|\nabla n|_{\mathrm{tPBE}} - |\nabla n|_{\mathrm{PBE}})/|\nabla n|_{\mathrm{tPBE}}$.
    (e) Electronic density profiles along $x$, $y$, and $z$ directions comparing relative differences of PBE (blue), and thermal PBE (red) with respect to PIMC reference data.}
\label{fig:elec_dens} 
\end{figure*}

In addition to thermodynamic quantities, we also examine the electronic charge density and electron localization function (ELF)~\cite{10.1063/1.458517}, which probe the XC functional in real space. We analyze a representative snapshot of hydrogen at $r\s=4$, $\theta=1$ ($T_F = 36{,}000~\mathrm{K}$), consisting of 32 atoms in a 
cubic cell (Fig.~\ref{fig:elec_dens}a). The electronic charge density and ELF are evaluated on $140 \times 140 \times 140$ and $90 \times 90 \times 90$ real--space grids, respectively.
Figure~\ref{fig:elec_dens}b shows relative differences in electronic density between thermal PBE and PBE along central slices of the [001], [010], and [100] planes, with red indicating increased density and blue indicating decreased density due to thermal effects. The changes are spatially smooth and small in magnitude, reflecting modest thermal redistribution of electrons. In contrast, Fig.~\ref{fig:elec_dens}c reveals that the ELF exhibits pronounced, highly localized differences reaching approximately $\pm15\%$. Thermal XC effects consistently enhance electron localization near protons while reducing it in interstitial regions. This pattern, observed across all crystallographic planes, indicates that finite--temperature XC corrections primarily redistribute electronic localization rather than uniformly smearing the density. 
Figure~\ref{fig:elec_dens}d shows the relative differences in the density--gradient magnitude, $(|\nabla n|_{\text{tPBE}} - |\nabla n|_{\text{PBE}})/|\nabla n|_{\text{tPBE}}$, along the [001], [010], and [100] planes. The spatial distribution reveals that finite--temperature XC effects modify gradient steepness in a highly nonuniform manner. Positive differences (red) near ionic positions indicate that ground--state PBE underestimates the steepness of density profiles in localization regions, while negative differences (blue) in interionic areas show that PBE overestimates gradient sharpness where thermal effects promote delocalization. This enhanced sensitivity of the density gradient to thermal XC corrections demonstrates that density--gradient analysis provides a more discriminating probe of finite--temperature electronic structure than density alone.

To provide quantitative validation, Fig.~\ref{fig:elec_dens}e compares one-dimensional density profiles from thermal PBE and PBE against PIMC reference data along the principal Cartesian directions~\cite{doi:10.1021/acs.jctc.3c00934}, where $\Delta n = (n_{\mathrm{DFT}} - n_{\mathrm{PIMC}}) / n_{\mathrm{PIMC}} \times 100$ denotes the percentage deviation of the DFT electronic density from the PIMC reference. Thermal PBE consistently shows slightly smaller deviations from PIMC than ground-state PBE across all three directions, with mean absolute deviations of 1.9\%, 1.5\%, and 1.2\% along $x$, $y$, and $z$ respectively, compared to 2.0\%, 1.6\%, and 1.3\% for PBE. 

These results underscore the importance of thermal XC corrections for interpreting bonding and electronic properties in warm dense hydrogen, particularly under conditions relevant to planetary interiors and inertial confinement fusion. We provide additional results including the temperature dependence of total and XC free energies of thermal PBE, as well as further details on the temperature dependence of charge density, ELF, and density gradients in Appendix~\ref{appendix.additional.results}.

\section{Conclusion}
We have implemented the thermal PBE functional -- an explicitly temperature--dependent GGA for the XC free energy in plane–wave electronic-structure codes. We assessed its performance systematically for warm dense matter.
Benchmarks against PIMC calculations show that thermal PBE provides a substantially improved description of warm dense hydrogen compared with ground--state LDA and PBE, as well as thermal LDA, yielding accurate energies, forces, pressures, and electronic densities at negligible additional computational cost. A systematic comparison with the nonempirical thermal GGA of Karasiev, Dufty, and Trickey (KDT16)~\cite{karasiev2018nonempirical} would be a natural extension of the present work and is left for future study. Overall, our results establish thermal PBE as a practical and reliable semilocal functional for first--principles simulations in the warm dense regime. The present implementation further enables the generation of finite--temperature DFT datasets for training machine--learning interatomic potentials applicable to extreme conditions~\cite{PhysRevB.106.L180101,Nikolov2024PNAS,kumar2024onthefly,PhysRevResearch.5.033162,Lindsey2025}. Beyond static properties, it also provides a foundation for investigations of excited--state and transport phenomena~\cite{PhysRevB.83.235120,PhysRevE.105.065204,PhysRevLett.116.115004,PhysRevB.103.125118,PhysRevB.107.115131,doi:10.1063/5.0138955,doi:10.1063/5.0135729,Moldabekov2023JPCL}, where explicit thermal XC effects are expected to play a critical role.

\begin{acknowledgements}
We thank J.~K. Dewhurst for the hospitality at the Max--Planck--Institut f\"{u}r Mikrostrukturphysik, Halle, and for fruitful discussions. K.R. acknowledges funding from the HPC Gateway project (Grant Agreement No.~1077561005) through the HHPAKT fund. This work was supported by the Center for Advanced Systems Understanding (CASUS), which is financed by Germany’s Federal Ministry of Research, Technology and Space (BMFTR) and by the Saxon State government out of the State budget approved by the Saxon State Parliament. This work has received funding from the European Research Council (ERC) under the European Union Horizon 2022 research and innovation program (Grant agreement No. 101076233, \textit{PREXTREME}). Views and opinions expressed are, however, those of the authors only and do not necessarily reflect those of the European Union or the European Research Council Executive Agency. Neither the European Union nor the granting authority can be held responsible for them. T.D. gratefully acknowledges funding from the Deutsche Forschungsgemeinschaft (DFG) via project DO 2670/1-1. K.B. acknowledges support from NSF grant No. DMR-2427903. 

Computations were performed on the Barnard Cluster at the Center for Information Services and High--Performance Computing (ZIH) at Technische Universit\"at Dresden, at the Norddeutscher Verbund f\"ur Hoch-- und H\"ochstleistungsrechnen (HLRN) under grant mvp00024, on the clusters Hemera and Rosi at Helmholtz--Zentrum Dresden--Rossendorf (HZDR), and on the HoreKa supercomputer at Karlsruher Institut f\"ur Technologie (KIT) funded by the Ministry of Science, Research and the Arts Baden--W\"urttemberg and by the Federal Ministry of Education and Research.
\end{acknowledgements}

\clearpage

\section*{Appendix}
\appendix

\section{Thermal enhancement prefactor}
\label{appendix.thermal.prefactor}

\begin{figure}[!htbp]
\centering
\includegraphics[width=1.0\columnwidth]{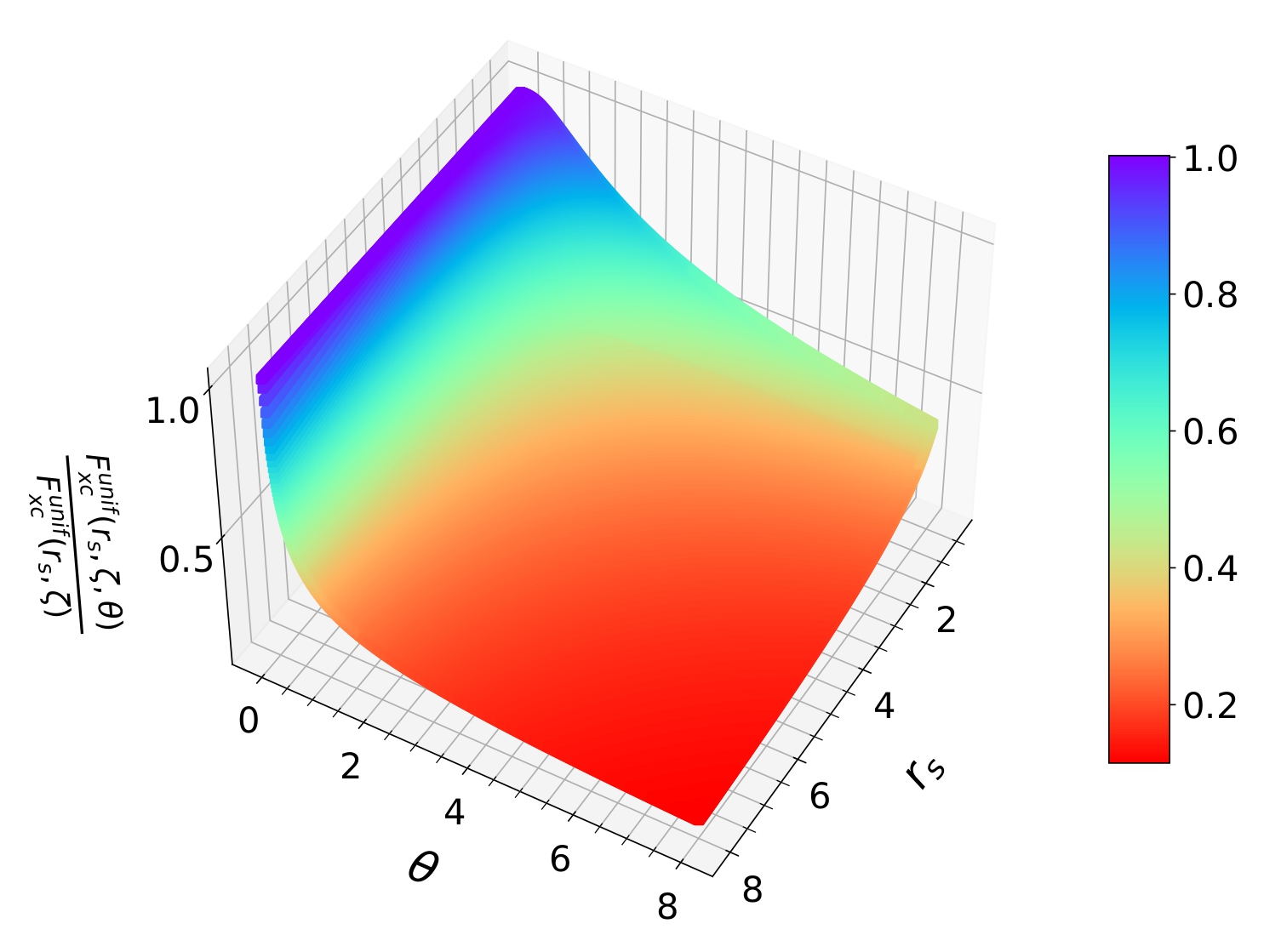}
\caption{\raggedright Thermal prefactor $F\xc^{\unif}(r\s, \zeta, \theta)/F\xc^{\unif}(r\s, \zeta)$ in Eq.~\ref{enhancement_factor_eqn} as functions of density ($r\s$) and reduced temperature ($\theta=T/T\F$) for the spin--unpolarized ($\zeta=0$) uniform electron gas.}
\label{ef}
\end{figure}

In the following, we investigate how changes in density and temperature influence the thermal PBE enhancement factor. For the enhancement factors of the thermal and ground--state uniform electron gas, we use the parameterization by Groth \textit{et al.}~\cite{PhysRevLett.119.135001}. We examine the impact of the thermal prefactor (Eq.~\ref{enhancement_factor_eqn}), $F\xc^{\unif}(r\s, \zeta, \theta)/F\xc^{\unif}(r\s, \zeta)$. Figure~\ref{ef} illustrates this prefactor for spin--unpolarized uniform electron gas ($\zeta=0$) at reduced temperature ($\theta=0-8$) and densities ($r\s=1-8$).  At high densities (small $r\s$), thermal XC effects become significantly more pronounced at higher temperatures. However, as density decreases, the effects of finite--temperature XC are pronounced even at relatively lower temperatures~\cite{PhysRevE.93.063207,PhysRevB.101.195129}.

\section{Validating the implementation of the thermal PBE functional}
\label{appendix.implementation}
To validate our implementation, we perform calculations for the uniform electron gas across a range of temperatures and densities using an in--house version of the full--potential linearized augmented--plane--wave (FP--LAPW) method~\cite{singh2006planewaves} implemented in the Elk code~\cite{elk}. In these calculations, we employ $16\times16\times16$ and  $32\times32\times32$ \emph{k}--point grids with the occupation of the highest band constrained to $\sim10^{-7}$.

\begin{figure*}[!htbp]
\centering
\includegraphics[width=1.7\columnwidth]{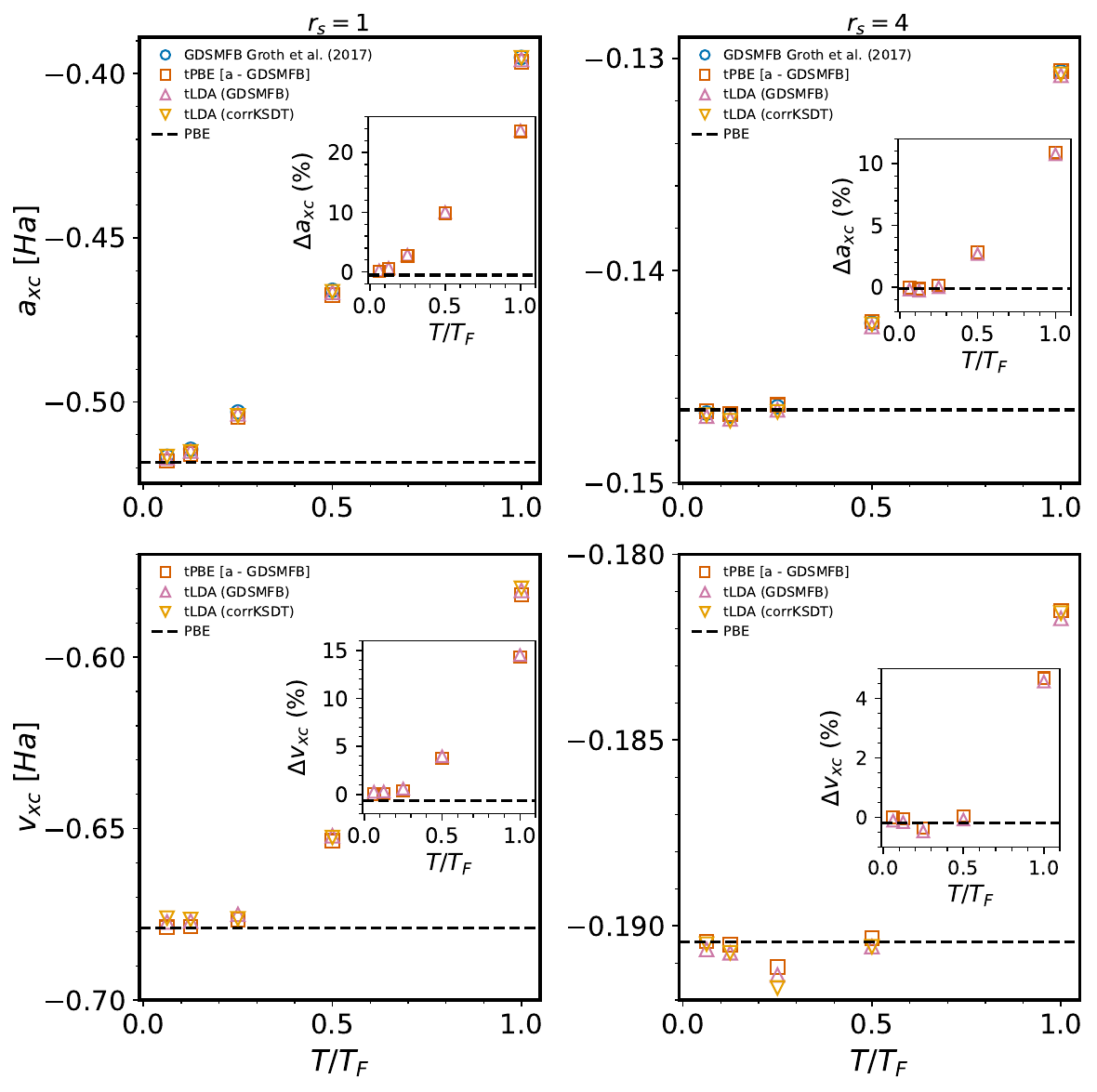}
\caption{\raggedright Exchange--correlation free energy per particle ($a_\textnormal{xc}$) and the XC potential energy per particle ($v_\textnormal{xc}$) for the spin--unpolarized uniform electron gas as a function of reduced temperature at densities ($r\s$=1 and $r\s$=4) in comparison with the GDSMFB parameterization of the uniform electron gas by Groth \textit{et al.}~\cite{PhysRevLett.119.135001}. The thermal LDA functional is evaluated using the GDSMFB parameterization, as is the thermal prefactor (Eq.~\ref{enhancement_factor_eqn}) of thermal PBE. In addition, we include the thermal LDA functional evaluated using the corrKSDT parameterization of Karasiev \textit{et al.}~\cite{karasiev2018nonempirical}. The inset plots show the relative differences between PBE, thermal LDA, and thermal PBE highlighting the importance of the explicit temperature dependence in the thermal prefactor as temperature increases.}
\label{xc_HEG}
\end{figure*}

\begin{figure*}[!htbp]
\centering
\includegraphics[width=1.7\columnwidth]{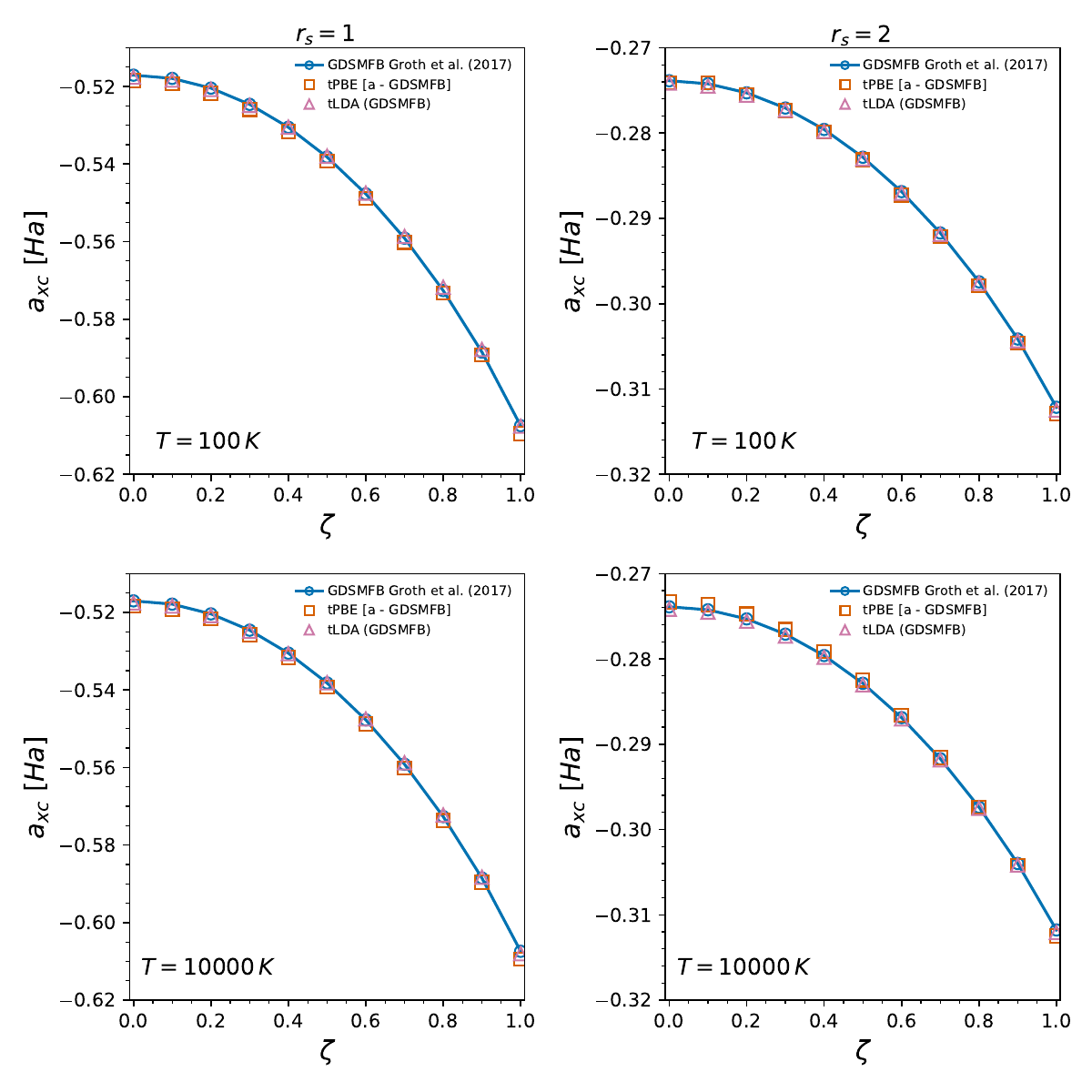}
\caption{\raggedright Exchange--correlation free energy per particle ($a_\textnormal{xc}$) for the uniform electron gas as a function of spin--polarization ($\zeta$) at densities $r\s$=1 and $r\s$=2, shown for temperatures  $T$=100~K (top row) and   $T$=10,000~K (bottom row). Results are compared with the GDSMFB parameterization of Groth \textit{et al.}~\cite{PhysRevLett.119.135001}. The thermal LDA functional is evaluated using the GDSMFB parameterization, as is the thermal prefactor (Eq.~\ref{enhancement_factor_eqn}) of thermal PBE.}
\label{xc_HEG_pol}
\end{figure*}

Figure~\ref{xc_HEG} displays the results of this assessment. We show the XC free energy per particle ($a\xc$) and the XC potential energy per particle ($v\xc$) for a range of reduced temperatures at densities $r\s=1$ and $r\s=4$. The reference point is the parameterization of the exact PIMC data of Groth \textit{et al.}~\cite{PhysRevLett.119.135001}. We include results for the thermal PBE, thermal LDA, and PBE functionals. For thermal LDA and thermal PBE, the GDSMFB parameterization~\cite{PhysRevLett.119.135001} of the uniform electron gas is employed. We additionally include the thermal LDA evaluated using the corrKSDT parameterization of Karasiev 
\textit{et al.}~\cite{karasiev2018nonempirical} as a further reference. Importantly, Fig.~\ref{xc_HEG} shows that thermal PBE is correctly reduced to both the thermal LDA (in the uniform--density limit) and PBE at zero temperature. Moreover, thermal PBE shows excellent overall agreement with the PIMC reference data (GDSMFB), as expected. The insets further show the relative differences between thermal PBE and PBE, highlighting the growing deviation of thermal PBE from ground--state PBE at elevated temperatures. This agreement confirms the correctness and numerical stability of our implementation.

Although all hydrogen calculations employ $\zeta=0$, we additionally verify 
the spin--polarization dependence of the tPBE implementation for the uniform electron gas across $\zeta\in[0,1]$ for completeness. In Fig.~\ref{xc_HEG_pol} we examine the spin--polarization dependence of the XC free energy per particle ($a_\textnormal{xc}$) at densities $r\s=1$ and $r\s=2$, and temperatures $T$=100~K and $T$=10,000~K. The results are compared against the GDSMFB parameterization of Groth \textit{et al.}~\cite{PhysRevLett.119.135001}, which serves as the reference. As expected, the XC free energy exhibits a smooth and monotonic dependence on the spin--polarization ($\zeta$), with the fully spin--polarized limit ($\zeta=1$) yielding more negative values than the unpolarized case ($\zeta=0$), consistent with the dominant role of exchange effects. The tLDA, constructed directly from the GDSMFB parameterization, reproduces the reference results by construction. Importantly, the tPBE functional preserves the correct spin--polarization trends across all densities and temperatures considered. 

\section{Comparison of thermal and ground--state functionals}
\label{appendix.additional.results}

\begin{figure}[!htbp]
\centering
\includegraphics[width=1.0\columnwidth]{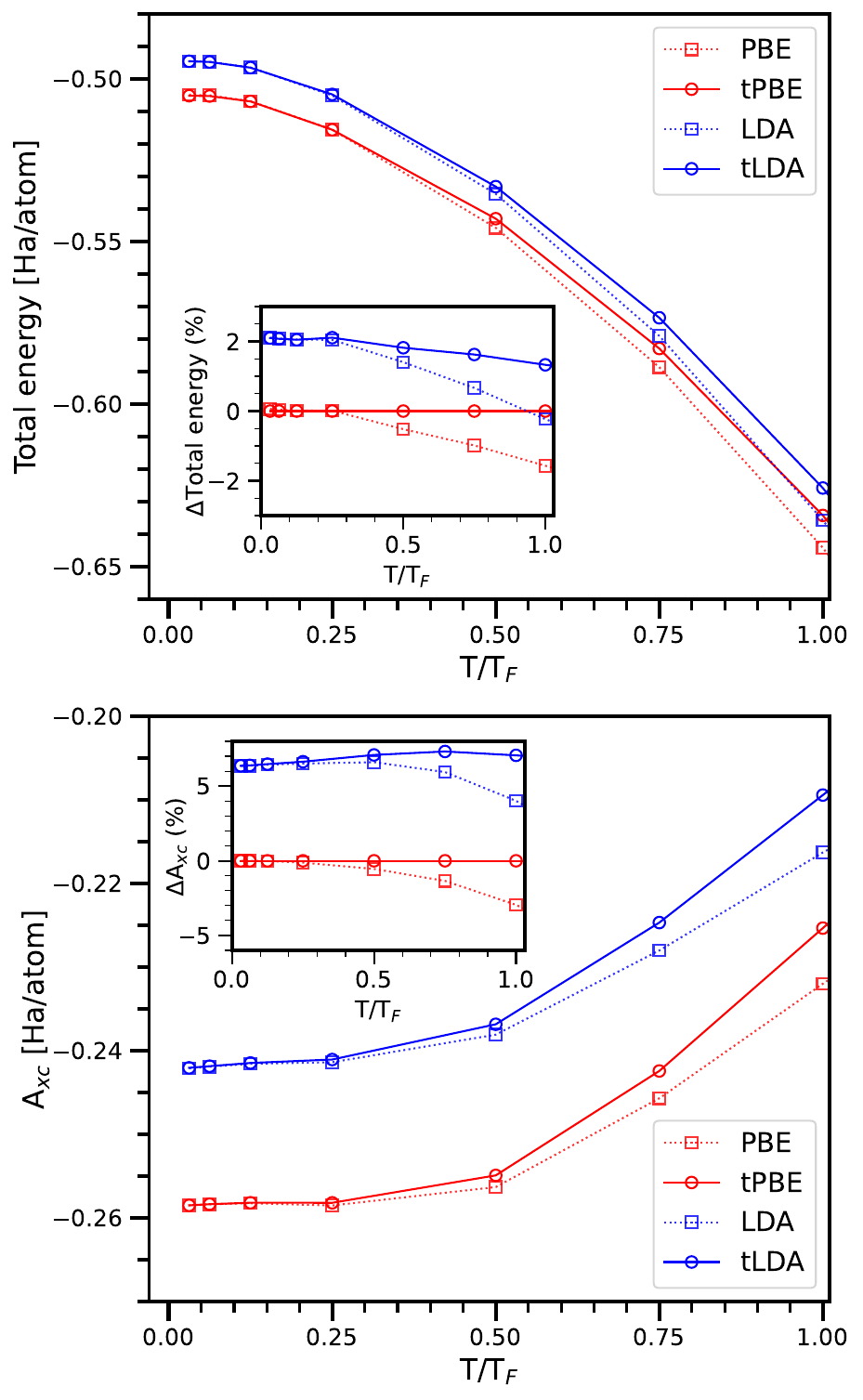}
\caption{\raggedright Total energy and XC free energy of hydrogen at $r\s$=4 as a function of reduced temperature ($\theta$=$T/T\F$) evaluated using ground--state (PBE, LDA) and thermal functionals (thermal PBE, thermal LDA). The inset plots show the relative difference in total energy and XC free energy between thermal PBE and other functionals.}
\label{hydrogen_rs4_dft_functionals}
\end{figure}

Figure~\ref{hydrogen_rs4_dft_functionals} compares the total energy and XC free energy of hydrogen at fixed density ($r\s$=4) across LDA, PBE, thermal LDA, and thermal PBE functionals as computed with the Elk code. In the top plot (total free energy), both thermal functionals correctly reduce to their ground--state counterparts at zero temperature. As the temperature increases, the difference between the thermal and ground--state functionals grows. The inset highlights relative differences with respect to thermal PBE, remaining within a few percent. The bottom plot depicts a similar comparison, but just for the XC free energy, revealing an analogous trend and a relative difference reaching approximately 3\% near $T\F$. 

\begin{figure}[!htbp]
\centering
\includegraphics[width=1.0\columnwidth]{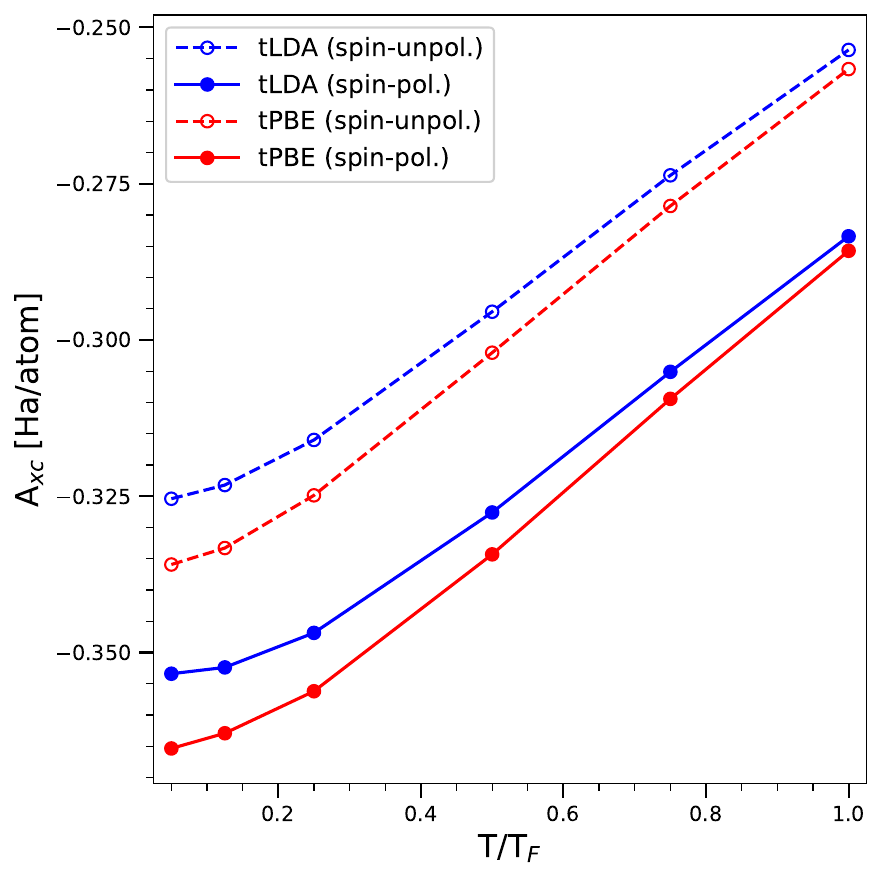}
\caption{\raggedright XC free energy of hydrogen at $r\s$=2 as a function of reduced temperature ($\theta$=$T/T\F$) evaluated using spin--unpolarized ($\zeta=0$) and spin--polarized ($\zeta=1$) thermal functionals (thermal PBE and thermal LDA).}
\label{hydrogen_rs2_spinpol}
\end{figure}

Figure~\ref{hydrogen_rs2_spinpol} compares the XC free energy of hydrogen at fixed density ($r\s$=2) across thermal LDA and thermal PBE functionals using spin--unpolarized ($\zeta = 0$) and spin--polarized ($\zeta = 1$) configurations. As expected, the spin--polarized XC free energy is more 
negative than the spin--unpolarized case across the entire temperature range, consistent with the dominant role of exchange effects. Importantly, thermal PBE correctly reproduces the spin--polarization dependence of thermal LDA across all temperatures considered. We note, however, that the spin--polarized ($\zeta = 1$) results are shown here solely for the purpose of validating the spin--polarization dependence of the thermal PBE functional, and are not physically appropriate for warm dense hydrogen. As discussed in Section~\ref{ref_results}, the spin--unpolarized configuration ($\zeta = 0$) is the physically correct choice for hydrogen in the WDM regime.

\begin{figure*}[!htbp]
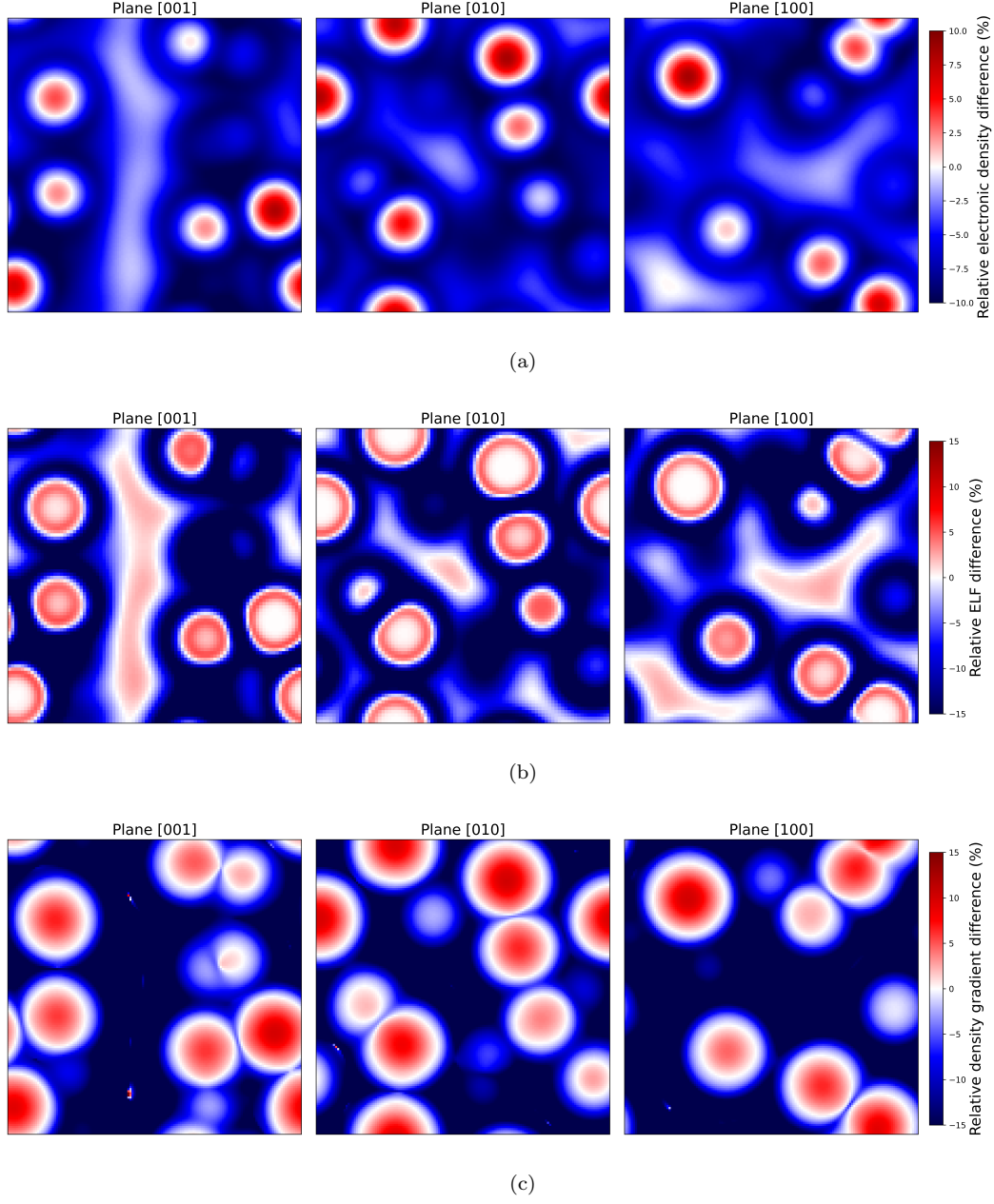

\centering
    \subfigure[]{\includegraphics[width=1.7\columnwidth]{charge_density_difference_planes_percentage_tpbe_tlda.png}}\\[0.5em]
    \subfigure[]{\includegraphics[width=1.7\columnwidth]{ELF_difference_planes_percentage_tpbe_tlda.png}}\\[0.5em]
    \subfigure[]{\includegraphics[width=1.7\columnwidth]{density_gradient_difference_planes_percentage_tpbe_tlda.png}}
\caption{\raggedright Comparison of thermal PBE and thermal LDA functionals across [001], [010], and [100] planes. (a) Charge density difference, (b) electron localization function (ELF) difference, and (c) density gradient difference. The percentages indicate the relative deviation between thermal PBE and thermal LDA.}
\label{fig:supp_tlda}
\end{figure*}

While Figs.~\ref{fig:elec_dens_b}, \ref{fig:elec_dens_c}, and \ref{fig:elec_dens_d} assess the impact of explicit thermal XC effects, we complement this analysis by examining density--gradient corrections. To this end, we analyze differences in the spatial distribution of electronic properties between thermal PBE and thermal LDA for hydrogen at $r\s=4$ and $\theta=1$, as shown in Fig.~\ref{fig:supp_tlda}. Specifically, we show the relative differences between thermal PBE and thermal LDA along the [001], [010], and [100] crystallographic planes, thereby isolating the contribution of density--gradient corrections in thermal PBE.

Panel (a) displays the charge density difference, $(n_{\text{tPBE}} - n_{\text{tLDA}})/n_{\text{tPBE}}$, revealing that density--gradient corrections in thermal PBE lead to modest but systematic redistribution of electronic density compared to thermal LDA. The differences are spatially smooth and exhibit characteristic patterns around ionic positions. Panel (b) presents the ELF difference, $(\text{ELF}_{\text{tPBE}} - \text{ELF}_{\text{tLDA}})/\text{ELF}_{\text{tPBE}}$, which shows more pronounced variations reaching approximately $\pm 10\%$. This indicates that the inclusion of density gradients substantially affects electron localization, particularly in the vicinity of protons and in interstitial regions. Panel (c) depicts the density--gradient magnitude difference, $(|\nabla n|_{\text{tPBE}} - |\nabla n|_{\text{tLDA}})/|\nabla n|_{\text{tPBE}}$, illustrating the relative sensitivity of the gradient of the density due to the density--gradient correction in thermal PBE.

Comparing Figs.~\ref{fig:elec_dens_b}, \ref{fig:elec_dens_c}, and \ref{fig:elec_dens_d} with Fig.~\ref{fig:supp_tlda} indicates that the impact of explicit thermal XC effects (thermal PBE versus ground-state PBE) is at least comparable in magnitude to that of density--gradient corrections (thermal PBE versus thermal LDA). This supports the conclusion that the thermal PBE functional incorporates both dependencies, each of which plays an important role in the warm dense regime.

\begin{figure*}[!htbp]
\centering
\includegraphics[width=1.7\columnwidth]{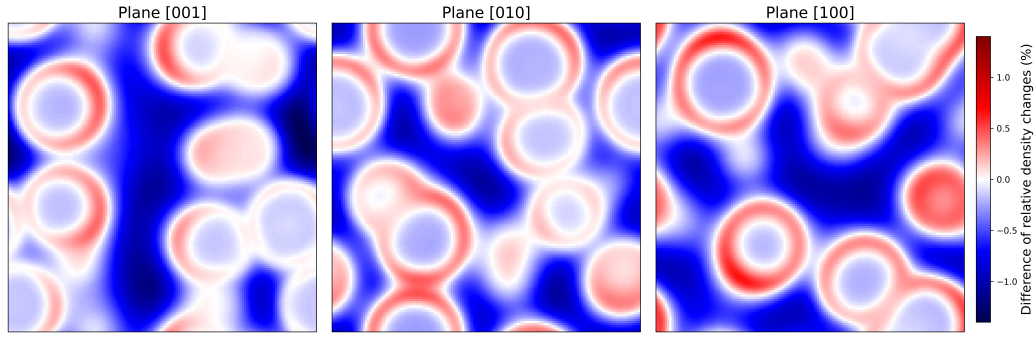}
\caption{\raggedright Difference between the relative electronic density changes across [001], [010], and [100] planes. Positive values (red) indicate regions where the relative density difference (thermal PBE vs.\ PBE) exceeds that of (thermal LDA vs.\ LDA); negative values (blue) indicate the opposite.}
\label{fig:supp_rel_rel_elec_dens}
\end{figure*}

To elucidate the distinct effects of thermal and XC functionals on the electronic density, we analyze the difference between the relative electronic density changes induced by two pairs of functionals: (i) thermal PBE versus PBE, and (ii) thermal LDA versus LDA as shown in Fig.~\ref{fig:supp_rel_rel_elec_dens}. We define the relative density change as

\begin{equation}
    \Delta_{\mathrm{rel}}^{X,Y} = 100 \times \frac{n_X - n_Y}{n_X},
\end{equation}
where $n_X$ and $n_Y$ denote the electronic densities obtained from functionals $X$ and $Y$, respectively. We then consider the difference between the relative changes for the two functional pairs:

\begin{equation}
    \delta\Delta_{\mathrm{rel}} = \Delta_{\mathrm{rel}}^{\mathrm{tPBE},\mathrm{PBE}} - \Delta_{\mathrm{rel}}^{\mathrm{tLDA},\mathrm{LDA}}.
\end{equation}

This quantity, $\delta\Delta_{\mathrm{rel}}$, highlights spatial regions where the impact of the thermal correction differs between the PBE and LDA families. Positive values of $\delta\Delta_{\mathrm{rel}}$ indicate that the relative change in density upon going from PBE to thermal PBE is greater than that from LDA to thermal LDA at a given spatial point, while negative values indicate the opposite. The spatial distribution of $\delta\Delta_{\mathrm{rel}}$ provides insight into the functional sensitivity of the electronic structure. Enhanced positive values near atomic centers or in bonding regions suggest that thermal PBE introduces a more pronounced correction relative to thermal LDA in those areas, whereas negative values indicate a stronger effect for thermal LDA. Regions where $\delta\Delta_{\mathrm{rel}} \approx 0$ indicate that both functional families yield similar relative changes upon inclusion of thermal effects.

\clearpage
\bibliography{bibliography}

\end{document}